\documentclass[12pt]{amsart}
\usepackage{amsmath}
\usepackage{amssymb}
\usepackage{latexsym}

\newcommand{\Zint}{{\bf Z}}    
     
\newtheorem{prop}{\bf Proposition}[section]
\newtheorem{thm}[prop]{\bf Theorem}

\newtheorem{exa}{\bf Example}

\begin{document}

\title{On the Inozemtsev model}
\author{Kouichi TAKEMURA}
\thanks{Supported by the Grant-in-Aid for Scientific Research (No. 15740108) from the Japan Society for the Promotion of Science}
\email{takemura@yokohama-cu.ac.jp}      
\address{Department of Mathematical Sciences, Yokohama City University, 22-2 Seto, Kanazawa-ku, Yokohama 236-0027, JAPAN}
%

\maketitle

\begin{abstract}
The $BC_N$ Inozemtsev model is investigated.
Finite-dimensional spaces which are invariant under the action of the Hamiltonian of the $BC_N$ Inozemtsev model are introduced and it is shown that commuting operators of conserved quantities also preserve the finite-dimensional spaces.
The $BC_2$ Inozemtsev model is studied in more detail.
\end{abstract}

\section{Introduction}

The $BC_N$ Inozemtsev model \cite{Ino} is a system of quantum mechanics with $N$-particles, which is a generalization of the Calogero-Moser-Sutherland model or the Olshanetsky-Perelomov model \cite{OP}. The Hamiltonian is given by
\begin{eqnarray}
& H=-\sum_{j=1}^N\frac{\partial ^2}{\partial x_j^2}+2l(l+1)\sum_{1\leq j<k\leq N} \left( \wp (x_j-x_k) +\wp (x_j +x_k) \right) \label{InoHam} & \\
& \; \; \; \; \; + \sum_{j=1}^N \sum _{i=0}^3 l_i(l_i+1) \wp(x_j +\omega_i), \nonumber & 
\end{eqnarray}
where $\wp (x)$ is the Weierstrass $\wp$-function with periods $(1,\tau )$, $\omega _0=0$, $\omega_1=1/2$, $\omega_2=-(\tau+1)/2$, $\omega_3=\tau/2$ are half periods, and $l$ and $l_i$ $(i=0,1,2,3)$ are coupling constants.

It is known that the $BC_N$ Inozemtsev model is quantum-completely integrable.
Here quantum-complete integrability means that there exist $N$ algebraically independent mutually commuting operators (higher commuting Hamiltonians) which commute with the Hamiltonian of the model. It is a quantum version of Liouville's integrability.
Note that the $BC_N$ Inozemtsev model is a universal completely integrable model of quantum mechanics with $B_N$ symmetry, which follows from the classification due to Ochiai, Oshima and Sekiguchi \cite{OOS,OS}.

In this paper we are going to investigate eigenvalues and eigenstates of the $BC_N$ Inozemtsev model especially for the case $N=2$. By the trigonometric limit $p =\exp(\mbox{i}\pi \tau) \rightarrow 0$, the Hamiltonian $H$ of the $BC_N$ Inozemtsev model tends to the Hamiltonian of the $BC_N$ Calogero-Moser-Sutherland model. 
More precisely, if $p \rightarrow 0$, then 
$H \rightarrow H_{T} +C_T$,
where
\begin{eqnarray}
& H_T = -\sum_{j=1}^N\frac{\partial ^2}{\partial x_j^2} +\sum_{1\leq j<k\leq N} \left( \frac{2\pi ^2 l(l+1)}{\sin ^2\pi (x_j-x_k)} +\frac{2\pi ^2l(l+1)}{\sin ^2 \pi (x_j +x_k)} \right) & \label{trigH} \\
& \; \; \; \; \; + \sum_{j=1}^N \left( \frac{\pi^2 l_0(l_0+1)}{\sin ^2 \pi x_j} + \frac{\pi^2 l_1(l_1+1)}{\cos ^2 \pi x_j} \right) , & \nonumber \\
& C_T= -\frac{N(N-1) \pi ^2}{3}l(l+1)-\frac{N\pi ^2}{3}\sum_{i=0}^3 l_i (l_i +1) . & \nonumber
\end{eqnarray} 
The operator $H_T$ is the Hamiltonian of the $BC_N$ trigonometric Calogero-Moser-Sutherland model.
Eigenvalues and eigenstates of the $BC_N$ Calogero-Moser-Sutherland model are known and the eigenstates are written by use of $BC_N$ Jacobi polynomials.
In contrast to this model it is much more difficult to find eigenvalues and eigenstates of the Inozemtsev model.

A possible approach to finding eigenvalues and eigenstates of the Inozemtsev model is to apply perturbation from the Calogero-Moser-Sutherland model.
Based on the eigenstates for the case $p=0$ we can obtain eigenvalues and eigenstates of the $BC_N$ Inozemtsev model $(p\neq 0)$ as formal power series in $p$.
Generally speaking, convergence of the formal power series obtained by perturbation is not guaranteed a priori, but for the case of the $BC_N$ Inozemtsev model the convergence radius of the formal power series in $p$ is shown to be non-zero (\cite{KT,Tak2,THI}) and it is seen that this perturbation is holomorphic.

Another approach is to use quasi-exact solvability, which is a main theme of this paper.
It is shown that, if the coupling constants $l$, $l_0$, $l_1$, $l_2$, $l_3$ satisfy some equation, the Hamiltonian $H$ (see (\ref{InoHam})) and the commuting operators of conserved quantities preserve a finite-dimensional space of doubly-periodic functions \cite{FGGRZ2,Takq}.
On the finite dimensional space, eigenvalues are calculated by solving the characteristic equation, and eigenfunctions are obtained by solving linear equations.
In this sense, part of the eigenvalues and eigenfunctions are obtained exactly and this is why the phrase ``quasi-exact solvability'' is used (see also \cite{Tur}).

This paper is organized as follows. 
In section \ref{sec:2}, finite-dimensional vector spaces which are related to quasi-exact solvability are introduced. In section \ref{sec:3}, it is seen that higher commuting operators also preserve the finite-dimensional invariant spaces. 
In section \ref{sec:4}, the $BC_2$ Inozemtsev model is investigated in more detail and in section \ref{sec:5}, we comment on the results.

\section{Invariant space} \label{sec:2}

We describe finite-dimensional spaces which are related to quasi-exact solvability.

We introduce the numbers $a$, $b_i$ $(i=0,1,2,3)$. Let $a$, $b_i$ $(i=0,1,2,3)$ be numbers which satisfy $a\in \{ -l, l+1 \}$ and $b_i \in \{-l_i/2, (l_i+1)/2 \}$ $(i=0,1,2,3)$.
Set 
\begin{eqnarray}
& z_j=\wp (x_j), \; \; \; 1\leq j\leq N, & \\
& \Phi(z)=\prod_{1\leq j<k\leq N} (z_j-z_k)^a \prod_{j=1}^N \prod _{i=1}^3 (z_j-e_i)^{b_i}, & \label{def:Phi}
\end{eqnarray}
where $e_i=\wp(\omega_i),$ $i=1,2,3$. We transform the Hamiltonian by
\begin{eqnarray}
& \widehat{H}= \Phi(z)^{-1} \circ H \circ  \Phi(z), \label{Hhat} &
\end{eqnarray}

Let $m_{m_1, m_2, \dots ,m_N}$ be the symmetric monomial defined by $$m_{m_1, m_2, \dots ,m_N}= \sum z_1^{n_1}z_2^{n_2}\dots z_N^{n_N} $$ where $n_1, n_2, \dots ,n_N$ runs over permutations of $m_1, m_2, \dots ,m_N$. Then we have the following proposition that describes a finite-dimensional invariant space:

\begin{prop} \cite{FGGRZ2, Takq} \label{prop:Hinv} 
Assume that $d=-((N-1)a+b_0+b_1+b_2+b_3)$ is a non-negative integer. Let $V^{\mbox{\scriptsize sym}}_d$ be the vector space spanned by symmetric monomials $m_{m_1, m_2, \dots ,m_N}$ such that $m_i \in \{ 0,1, \dots ,d\}$ for all $i$. 
Then $\widehat{H}  \cdot V_d^{\mbox{\rm \scriptsize sym}} \subset V_d^{\mbox{\rm \scriptsize sym}}$.
\end{prop}

\section{Commuting operators and invariant space} \label{sec:3} 

It is known that the $BC_N$ Inozemtsev model is completely integrable, i.e. there exist $N$ algebraically independent mutually commuting operators (higher commuting Hamiltonians) which commute with the Hamiltonian $H$.
In fact there exist operators $P_1, \dots , P_N$ such that 
\begin{equation}
[P_j, H]=0 , \; \; \; \; [P_j, P_k]=0 ,\label{Pjkcomm}
\end{equation}
for $1 \leq j,k \leq N$, the operator $P_k$ admits the expansion 
\begin{equation}
P_k= \sum_{1\leq j_1 <\dots <j_k \leq N} \left(\frac{\partial }{\partial x_{j_1}}\right)^{2} \dots \left(\frac{\partial }{\partial x_{j_k}}\right)^{2}+ \mbox{(lower terms)}
\end{equation}
and $P_1$ coincides with the Hamiltonian up to a constant. Note that in \cite{O} Oshima gave explicit forms of the commuting operators. Set
\begin{equation}
\widehat{P}_k= \Phi(z)^{-1} \circ P_k \circ  \Phi(z), \; \; \; (k=1, \dots ,N)
\label{def:widePk}
\end{equation}
where $\Phi(z)$ is defined in (\ref{def:Phi}). From (\ref{Pjkcomm}), we obtain
\begin{equation}
[\widehat{P}_j, \widehat{H}]=0 , \; \; \; \; [\widehat{P}_j, \widehat{P}_k]=0 \label{Phatjkcomm}
\end{equation}
for $1 \leq j,k \leq N$. Then it is shown that the operators $\widehat{P}_k$ $(k=1, \dots ,N)$ also preserve the space $ V_d^{\mbox{\rm \scriptsize sym}}$ which is related with quasi-exact solvability.

\begin{prop} \cite[Theorem 3.3]{Takq} \label{thm:Pinv} 
 Assume that $d=-((N-1)a+b_0+b_1+b_2+b_3)$ is a non-negative integer. Then $\widehat{P}_k  \cdot V_d^{\mbox{\rm \scriptsize sym}} \subset V_d^{\mbox{\rm \scriptsize sym}}$ for $k=1,2,\dots ,N$, where $V_d^{\mbox{\rm \scriptsize sym}}$ is the finite-dimensional space defined in proposition \ref{prop:Hinv}.
\end{prop}

In summary, we established by Proposition \ref{thm:Pinv} that higher commuting Hamiltonians also preserve the space related to quasi-exact solvablity.

\section{$BC_2$ Inozemtsev model} \label{sec:4}

In this section we investigate the $BC_2$ Inozemtsev model.

The transformed Hamiltonian $\widehat{H}$ (see (\ref{Hhat})) is written as

\begin{eqnarray}
& &  \widehat{H}=  - c(z_1) \left( \frac{\partial ^2}{\partial z_1^2} +\left( \frac{2a}{z_1-z_2}+b(z_1)\right) \frac{\partial}{\partial z_1} \right) \\
& & - c(z_2) \left( \frac{\partial ^2}{\partial z_2^2} +\left( \frac{2a}{z_2-z_1}+b(z_2)\right) \frac{\partial}{\partial z_2} \right) -4(a+\tilde{b})(a+\tilde{b}')(z_1+z_2)+2d_1+d_2 \nonumber 
\end{eqnarray}
where
\begin{eqnarray}
& c(z)=4(z-e_1)(z-e_2)(z-e_3) , \; \;  b(z)= \frac{2b_1+\frac{1}{2}}{z-e_1}+\frac{2b_2+\frac{1}{2}}{z-e_2}+\frac{2b_3+\frac{1}{2}}{z-e_3} , &  \\
& \tilde{b}=b_0 +b_1 +b_2 +b_3 , \; \;  \tilde{b}'=-b_0 +b_1 +b_2 +b_3 +\frac{1}{2} , & \nonumber \\
& d_1=4((b_1+b_2)^2e_3+(b_1+b_3)^2e_2+(b_2+b_3)^2e_1) ,& \nonumber \\
&  d_2= -8a(e_1b_1+e_2b_2+e_3b_3). & \nonumber
\end{eqnarray}
The transformed commuting operator $\widehat{P}_2$ is written as
\begin{eqnarray}
& & \widehat{P}_2=  c(z_1) c(z_2) \left[ \frac{\partial ^4}{\partial z_1^2 \partial z_2^2 } +\left( \frac{2a}{z_2-z_1}+b(z_2)\right) \frac{\partial ^3}{\partial z^2_1 \partial z_2} \right. \\
& & +\left( \frac{2a}{z_1-z_2}+b(z_1)\right) \frac{\partial ^3}{\partial z_1 \partial z^2_2} +q(z_1,z_2)  \frac{\partial ^2}{\partial z^2_1}+ q(z_2,z_1)  \frac{\partial ^2}{\partial z^2_2} \nonumber \\
& & +\left( \left( \frac{2a}{z_1-z_2} +b(z_1) \right) \left( \frac{2a}{z_2-z_1} +b(z_2) \right) +\frac{2a(a+1)}{(z_1-z_2)^2} \right) \frac{\partial ^2}{\partial z_1 \partial z_2}  \nonumber  \\
& & +\left( \frac{a(a+1)}{(z_2 -z_1)^2} \left( \frac{2(a-1)}{z_2-z_1} +b(z_2) \right) +\left( \frac{2a}{z_1-z_2} +b(z_1) \right)  q(z_1, z_2) \right) \frac{\partial }{\partial z_1} \nonumber  \\
& & + \left. \left( \frac{a(a+1)}{(z_1 -z_2)^2} \left( \frac{2(a-1)}{z_1-z_2} +b(z_1) \right) +\left( \frac{2a}{z_2-z_1} +b(z_2) \right)  q(z_2, z_1) \right)  \frac{\partial }{\partial z_2} \right] \nonumber \\
& & +4(a+\tilde{b})(a+\tilde{b}')(4\tilde{b} \tilde{b}'z_1z_2-d_1(z_1+z_2)) \nonumber 
\end{eqnarray}
where
\begin{equation}
 q(x,y)= \frac{a}{y-x}\left( \frac{a-1}{y-x} +b(y) \right) + \frac{4 \tilde{b} \tilde{b}' y -d_1}{c(y)} 
\end{equation}
Note that we normalize the operator $\widehat {P}_2$ as $\left. \left( \widehat {P}_2 \cdot 1 \right) \right| _{z_1=z_2=0} =0$.

From Proposition \ref{thm:Pinv} we obtain that if $d=-(a+b_0+b_1+b_2+b_3) \in \Zint _{\geq 0}$, then the operators $\widehat{H} (\sim \widehat{P}_1)$ and $\widehat{P}_2$ preserve the space $V_d^{\mbox{\rm \scriptsize sym}}$. Note that $V_d^{\mbox{\rm \scriptsize sym}} = \mbox{span} \{ z_1^{m_1} z_2^{m_2} +z_1^{m_2} z_1^{m_1} | 0\leq m_1 \leq m_2 \leq d \}$ and $\dim V_d^{\mbox{\rm \scriptsize sym}} =(d+1)(d+2)/2$.

Since the operator $\widehat{H}$ commutes with the operator $\widehat{P}_2$, it is natural to consider simultaneous diagonalization of the operators $\widehat{H}$ and $\widehat{P}_2$. It is observed that, if $d=1,2,3$, then the spectra of the operator $\widehat{H}$ do not have multiplicity in generic parameters. Therefore, it is reasonable to suppose that the operators $\widehat{H}$ and $\widehat{P}_2$ satisfy an algebraic relation on the space $V_d^{\mbox{\rm \scriptsize sym}}$. In fact, if $d=1$ or $2$, then we have the following theorem:

\begin{thm}
(i) If $d=-(a+b_0+b_1+b_2+b_3) =1$, then on the space $V_1^{\mbox{\rm \scriptsize sym}}$
\begin{equation}
\widehat{P}_2 =c_2 \widehat{H}^2 +c_1 \widehat{H} + c_0
\end{equation}
for some constants $c_2, c_1, c_0$.\\ 
(ii) If $d=-(a+b_0+b_1+b_2+b_3) =2$, then on the space $V_2^{\mbox{\rm \scriptsize sym}}$
\begin{equation}
\widehat{P}_2 ^2 +d_0 \widehat{P}_2 =c_4 \widehat{H}^4 +c_3 \widehat{H}^3 + c_2 \widehat{H}^2 +c_1 \widehat{H} + c_0
\end{equation}
for some constants $d_0, c_4, c_3, c_2, c_1, c_0$.
\end{thm}

\begin{exa}
(i) If $a=2$, $b_0=1$, $b_1=-1$, $b_2=-3$, $b_3=0$, then $d=-(a+b_0+b_1+b_2+b_3) =1$ and the operators $ \widehat{H}$ and $\widehat{P}_2 $ preserve the space $V_1^{\mbox{\rm \scriptsize sym}}$. We have 
\begin{equation}
\widehat{P}_2 =\frac{3}{2} \widehat{H}^2 +(40e_2-80e_3) \widehat{H} + 44e_2^2-500e_2e_3+380e_3^2
\end{equation}
on the space $V_1^{\mbox{\rm \scriptsize sym}}$. Note that we used a relation $e_1+e_2+e_3=0$.\\
(ii) If $a=-7$, $b_0=3$, $b_1=-1$, $b_2=2$, $b_3=1$, then $d=-(a+b_0+b_1+b_2+b_3) =2$ and the operators $ \widehat{H}$ and $\widehat{P}_2 $ preserve the space $V_2^{\mbox{\rm \scriptsize sym}}$. If $e_1=0$ and $e_2=-e_3$, then we have 
\begin{equation}
\widehat{P}_2 ^2 -8464e_3^2 \widehat{P}_2 =\frac{5}{2} \widehat{H}^4 +\frac{1040}{3}e_3 \widehat{H}^3 -1484e_3^2 \widehat{H}^2 -\frac{2279680}{3}e_3^3 \widehat{H} - 11061824e_3^4
\end{equation}
on the space $V_2^{\mbox{\rm \scriptsize sym}}$.
\end{exa}

\section{Comments} \label{sec:5}

In this paper, we see quasi-exact solvability not only for the Hamiltonian of the $BC_N$ Inozemtsev model but also for commuting operators of conserved quantities. In \cite{Takq} several relationships with other models based on quasi-exact solvability were clarified. It was seen that theta-type invariant spaces for the $BC_N$ Ruijsenaars model correspond to the spaces which are related to quasi-exact solvability for the $BC_N$ Inozemtsev model. Degeneration of the $BC_N$ Inozemtsev model (especially for its quasi-exact solvability) was also investigated.

In general the space $V_d^{\mbox{\rm \scriptsize sym}}$ is not a subspace of a suitable Hilbert space. The condition that $V_d^{\mbox{\rm \scriptsize sym}}$ is contained in the Hilbert space is written explicitly. For details see \cite{Takq}.

For the case $N=1$ it is known that finding eigenstates of the Hamiltonian is equivalent to solving the Heun equation. In this sense the $BC_N$ Inozemtsev model is a generalization of the Heun equation. For the case $N=1$ more results were obtained.
In \cite{Tak2} holomorphy of perturbation for the $BC_1$ Inozemtsev model was established and some inequalities of eigenvalues were shown. In \cite{Tak1,Tak3} the Bethe ansatz method for the $BC_1$ Inozemtzev model was proposed and some results on the finite-gap integration were established. Global monodromies of the Heun equation were calculated for the case when all coupling constants are integers, by using hyperelliptic integrals, and applications for analyzing eigenvalues with boundary conditions were obtained.

It is desirable to develop finite-gap integration and to obtain formulae of global monodromies for the $BC_N$ Inozemtzev model, although in my opinion this has not yet been done successfully. The result by Chalykh, Etingof and Oblomkov \cite{CEO} provides some hints as to how to approach this problem.

\end {document}